\documentclass[aps,pre,twocolumn,superscriptaddress]{revtex4}

\usepackage{amsfonts,amssymb,amsmath,latexsym,epsfig,wasysym}
\usepackage[sort&compress]{natbib}
\usepackage{amsmath}
\usepackage{xcolor}

\newcommand{\D}[0]{\mathcal{D}}

\begin{document}

\title{Spatial interactions in urban scaling laws}

\author{
Eduardo G. Altmann}
\affiliation{School of Mathematics and Statistics, The University of Sydney, Australia}
\affiliation{Centre for Complex Systems, The University of Sydney, Australia}


\begin{abstract}
Analyses of urban scaling laws assume that observations in different cities are independent of the existence of nearby cities. Here we introduce generative models and data-analysis methods that overcome this limitation by modelling explicitly the effect of interactions between individuals at different locations. Parameters that describe the scaling law and the spatial interactions are inferred from data simultaneously, allowing for rigorous (Bayesian) model comparison and overcoming the problem of defining the boundaries of urban regions. Results in five different datasets show that including spatial interactions typically leads to better models and a change in the exponent of the scaling law. Data and codes are provided in Ref.~\cite{github}.
 \end{abstract}

\maketitle

\section{Introduction}

One of the pillars of the study of cities as complex systems is the existence of statistical laws that apply ``universally'' to urban regions in different locations~\cite{Krugman2001,BattyBook,Barthelemy,Rybski2019}. Examples include the Zipf's law of city sizes, the gravitational law of population movement, and -- the focus of this paper -- scaling laws
\begin{equation}\label{eq.scaling}
  y \sim x^\beta,
\end{equation}
between observables $y$ and the population $x$ of cities.
All these laws have their origin in the first half of the XX century and continue to be investigated in increasingly rich datasets~\cite{Rozenfeld2011,Simini2012,Barnes2014}. In particular, the scaling law~(\ref{eq.scaling}) was discussed for the area of cities since the 1940s~\cite{Stewart1947}, can be viewed as a form of {\it increasing return to scale}~\cite{Krugman2001,Sarkar2018}, and has been the subject of many recent studies~\cite{Bettencourt2007a,Bettencourt2010,Bettencourt2013s,Louf2014,Arcaute2015,Ribeiro2017,Rybski2019}.

\begin{figure*}[!th]
  \includegraphics[width=0.9\columnwidth]{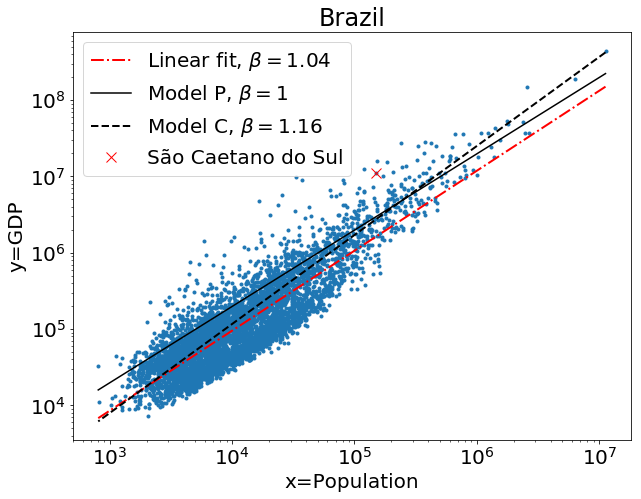}  \includegraphics[width=0.9\columnwidth]{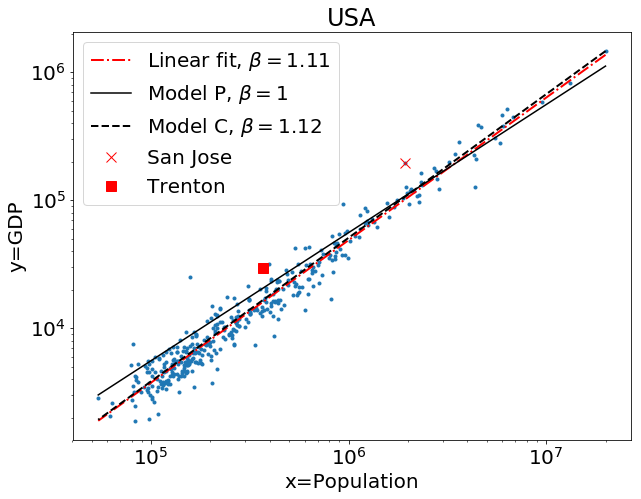}
\caption{Urban scaling laws~(\ref{eq.scaling}) based on different models. The GDP $y$ of different municipalities in Brazil (left) and metropolitan areas in USA (right) are shown as a function of their population $x$. The straight lines correspond to different models: linear fit of the data, the Per capita model (P) and the City model (C), see Eq.~(\ref{eq.ymodel}). The estimated scaling exponent $\beta$ of the different models are shown in the caption. Cities close to large urban areas are highlighted. }
\label{fig.1}
\end{figure*}

Originally, urban laws were seen as akin to the empirical laws of classical mechanics, the basis of a {\it sociophysics} theory~\cite{Stewart1947,Barnes2014}. A modern trace of this simplistic view is the fact that models and explanations of the origin of these laws are typically presented independently from the statistical analysis in support of their validity, e.g., the data analysis supporting~(\ref{eq.scaling}) is based on straight-line fits of $\log y$ vs. $\log x$ regardless of the explanation for its appearance. This undermines the statistical nature of the laws (evident from the large fluctuations) and is unable to select between the many alternative models that ``explain'' their origin (which often predict different fluctuations and can thus be tested).

The need for careful data-analysis methods to investigate statistical laws in complex systems has been extensively discussed for power-law distributions such as Zipf's law~\cite{Clauset2009,Gerlach2019,Corral2020}. Similar scrutinity is being applied to the methods used in scaling laws in urban systems~\cite{Shalizi2011,Louf2014,Arcaute2015,Leitao2016,Finance2019} and reveal the limitations of the traditional linear-fitting approach: it relies on several simplifying assumptions, it is unable to deal with $y=0$ in the data, it makes it difficult to compare to alternative models and to assess whether the scaling is indeed non-linear ($\beta \neq 1$), and it treats each city equally so that results are sensitive to cut-offs and fluctuations in the data of the many small cities. These limitations motivated us to introduce in Ref.~\cite{Leitao2016} a model of urban scaling that focuses on individuals instead of cities, effectively giving more weight to the largest cities. Fig.~\ref{fig.1} compares this and alternative fitting models for the dependence  of the Gross Domestic Product (GDP, $y$) on the population of cities ($x$) in two countries. 

A limitation that persists is that all data-analyses methods of scaling laws~(\ref{eq.scaling}) ignore the crucial element of any urban data: their spatial component~\cite{Krugman2001}. Linear fitting and all methods proposed in Ref.~\cite{Leitao2016} assume that observations in different cities are independent from each other and thus independent of their location. Not surprisingly, the scalings show spatially-correlated fluctuations~\cite{Bettencourt2010} and are sensitive to the definition of city boundaries~\cite{Louf2014,Arcaute2015}. For instance, in the results in Fig.~\ref{fig.1} (left panel) we highlight one of Brazil's municipalities (``S\~ao Caetano do Sul''-SP) that lies within Brazil's largest metropolitan area (around ``S\~ao Paulo''-SP). We see that the GDP of this municipality is much larger than expected by any of the models and it is natural to suspect that this is at least partially due to its proximity to other urban areas. This effect is enhanced by the fact that Brazil's data is aggregated according to administrative areas (municipalities), which often do not reflect connected urban regions. Still, the problem of defining appropriate urban areas is not trivial~\cite{Rozenfeld2011,Arcaute2015} and spatial proximity should play a role regardless of the chosen urban unit. In fact, Fig.~\ref{fig.1} (right panel) shows that in USA, where data is given for metropolitan areas, a similar effect appears (e.g., ``San-Jos\'e-Santa Clara''-CA close to ``San Francisco''-CA, or ``Trenton''-NJ between ``New York City''-NY and ``Philadelphia''-PA).

Here we propose a framework to investigate urban scaling~(\ref{eq.scaling}) that is based on generative models (Sec.~\ref{sec.model}), accounts for (spatial) interactions between different urban areas, and allows for rigorous statistical analyses (Sec.~\ref{sec.analysis}).  Results in 5 datasets from Brazil and USA show (Sec.~\ref{sec.results}) that, in most cases, models that accounts for spatial interactions provides a better description of the data and that the scaling exponent $\beta$ depends on the spatial scaling, in agreement with previous observations~\cite{Louf2014,Arcaute2015} of the dependence of $\beta$ on the urban unit.

\section{Model}\label{sec.model}

\subsection{Generative model}

The starting point of our model is the widespread interpretation that
Eq.~(\ref{eq.scaling}) reflects a change in people's efficiency (or consumption) depending on the amount of interactions available to them~\cite{Bettencourt2013s}.
Accordingly, we consider a generative process in which tokens (e.g. a patent,
a dollar of GDP, a piece of infrastructure) are assigned to (produced or consumed by) an individual person $j$ with probability $p(j)$.

Consider $j=1,...,M$ persons living in  $i=1,...,N$ cities, on which the population of the city $i$ is given by $x_i$ and $X=\sum_i^N x_i$.
A total of $Y \equiv \sum_i y_i$ tokens are (randomly) assigned to the $X$ persons. In the absence of any other information, this defines our first (null) model:

\begin{itemize}
\item[(P)] {\it Per-capita model:} all tokens $Y$ are distributed with equal probability to all persons $j$ as in a constant per-capita attribution, $p(j)=1/X$. In this case, the probability that a token is attributed to city $i$ is given by
  \begin{equation}\label{eq.P}
p(i) = \sum_j p(j) \delta(c(j)-i) = \frac{x_i}{X},
    \end{equation}
 where $c(j)$ is the city in which $j$ lives and $\delta(x)=1$ for $x=0$ (otherwise $\delta(x)=0$).

  \end{itemize}

This model corresponds to a linear (trivial) scaling law, $\beta=1$ in Eq.~(\ref{eq.scaling}). A super-linear $\beta>1$ (sub-linear $\beta<1$) scaling is obtained if a token is more likely to be assigned to someone living in a more (less) populous city.  In this spirit, in Ref.~\cite{Leitao2016} we assumed that the probability that a token is assigned to person $j$ depends on the population around $j$ as
\begin{equation}\label{eq.pj0}
  p(j) \sim x_{c(j)}^{\beta-1}.
\end{equation}
 Here we generalize this idea to account for spatial interactions between  $j$ and other individuals $j'$ that live in other cities (i.e., $c(j) \ne c(j')$) and therefore we write
\begin{equation}\label{eq.pj}
p(j) = \frac{A_j^{\beta-1}}{Z(\beta)},
\end{equation}
where $A_{j}$ is the total attractiveness due to all interactions of $j$ and $Z({\beta})$ is the normalization constant (i.e., $\sum_j^X p(j)=1$). If $\beta = 1$, the probability $p(j)$ is the same for all $j$ as in the per-capita model and we recover Eq.~(\ref{eq.P}). For $\beta>1$, $p(j)$ grows with the interactions $A_j$ in line with a super-linear scaling. For $\beta<1$, $p(j)$ decays with $A_j$ in line with a sub-linear scaling. 

The attractiveness of an individual $A_j$ certainly depends on a multitude of factors that could be included in the model, depending on data availability and research interest. Here, we focus on pairwise interactions $a_{j,j'}$ between individuals $j$ and $j'$ separated by a distance $d=d_{j,j'}$. We obtain $A_j$ as the total interaction of $j$ and all other individuals $j'$ by summing over all $j'$
\begin{equation}\label{eq.Aj}
A_j = \sum_{j'\neq j} a_{j,j'}(d_{j,j'}).
\end{equation}
The distance $d_{j,j'} \ge 0$ does not need to be a distance in a mathematical sense and, in practice, depends on the availability of data. Below we use the geographic (geodesic) distance between cities (another natural choice would be the commuting time).  The pairwise (spatial) interactions $a_{j,j'}$ is discussed below and will lead to three different specific models. 

\subsection{Spatial interactions}

In order to explore the formalism above we now consider simple dependencies of the pairwise interaction $a(d)$ on the distance $d \equiv d_{j,j'}$ between two  persons $j,j'$. In general, we are interested in functions $a(d)$ that monotonically decay with $d$ from $a(0)=1$ to $\lim_{d \rightarrow \infty} a(d)=0$\footnote{Choosing another value at $a(0)$ leads to the same results because of the normalization of $p(j)$ in Eq.~(\ref{eq.pj}).}. Our framework can be applied to any function $a(d)$ suitable to model spatial relationships, the analysis of data will reveal us which one is more suitable.

The simplest choice of $a(d)$ is

\begin{itemize}
\item[(C)] {\it City model:} 
\begin{equation}\label{eq.a0}
a_C(d) = \delta(d) = \left\{\begin{array}{l} 
1 \text{ if } d= 0 \text{ (or } c(j)=c(j')\text{)} \\
0 \text{ if }  d>0 \text{ (or } c(j) \neq c(j')\text{)}
\end{array}\right.,
\end{equation}
in which interactions occur only within the same city ($d=0$). From Eq.~(\ref{eq.Aj}) we get $A_j = x_{c(j)}$, i.e., $A_j$ and thus we recover the scaling law~(\ref{eq.scaling}) and Eq.~(\ref{eq.pj0}) (the model of Ref.~\cite{Leitao2016}, Sec. 4.2).
\end{itemize}

Spatial interactions beyond city limits can be incorporated using more general functions $a(d)$. Here we start this investigation with functions $a(d;\alpha)$ that depend on a single parameter $\alpha$ that is measured in the same units of $d$ (e.g., $km$) and sets a scale for spatial interactions such that $a(\alpha;\alpha) = 1/2$ (i.e., at a distance $d=\alpha$ the interactions decay to a factor $0.5$ of the interaction at the same city $d=0$). Furthermore, we wish to recover the choice~(\ref{eq.a0}) in the limit of small $\alpha$, i.e., $a(d) \rightarrow a_C(d)$ in Eq.~(\ref{eq.a0}) for $\alpha \rightarrow 0_+$. Two choices of $a(d;\alpha)$ that satisfy these properties (and also $a(0;\alpha) = 1$ and $\lim_{d \rightarrow \infty} a(d,\alpha)=0$ for any $\alpha$) are:

\begin{itemize}

  \item[(G)] {\it Gravitational model:}
\begin{equation}\label{eq.a}
a_G(d;\alpha_G) = \frac{1}{1+ \left(\frac{d}{\alpha_G}\right)^2},
\end{equation}
inspired by models of gravitational interactions (for large $d$ the interactions decay as $a \sim 1/d^2$, one can also replace the power $2$ by an additional parameter)~\cite{Stewart1947,Barthelemy,Ribeiro2017}.

\item[(E)] {\it Exponential model:}

  \begin{equation}\label{eq.aexp}
    a_E(d;\alpha_E) = e^{- d \ln(2) / \alpha_E}.
    \end{equation}

\end{itemize}

For $\alpha\rightarrow\infty$, the distances do not matter, everyone is equally linked to everyone else, and the $P$-model is retrieved. Altogether, the four models discussed above are summarized in Tab.~\ref{tab.models} and satisfy

$$ C \xleftarrow[\alpha\rightarrow 0]{}(G,E) \xrightarrow[\alpha\rightarrow \infty]{\beta \rightarrow 1} P .$$

\begin{table}[bt]
\begin{center}
  \begin{tabular}{|c c | c | c|}
    \hline
    \multicolumn{2}{|c|}{Model} & Attractiveness $a(d)$ & Parameters $\theta$\\
    \hline
    Per capita & P & - & - \\
    City & C & $\delta(d)$ & $\beta$ \\
    Gravitational & G & $1/(1+(d/\alpha)^2)$ & $\alpha,\beta$ \\
    Exponential & E & $e^{-d\ln 2/\alpha}$ & $\alpha,\beta$ \\
    \hline
  \end{tabular}\label{tab.models}
    \caption{The four models considered in this paper.}
\end{center}
\end{table}

\subsection{Likelihood}

We now discuss how the likelihood of our models can be computed from the data. We assume that $(x_i,y_i)$ data is available at locations $i=1,\ldots,N$. We denote the locations $i$ as city but we stress that this does not need to correspond to any urban definition of cities as the spatial interaction between different regions can be accounted explicitly in our models by choosing an appropriate function $a(d)$. We assume also that a measure of distance $d_{i,i'}$ between all pairs of cities is available (e.g., the geodesic distance between the centroid of the cities).

Besides their location (city), individuals are indistinguishable. Therefore, the probability $p(i)$ that a token is assigned to city $i$ is given by a sum of $p(j)$ over persons $j$ on city $i$ (i.e. $c(j)=i$), which contains exactly $x_{c(j)}\equiv x_i$ terms
\begin{eqnarray}
  p(i) &=& \sum_{j} p(j) \delta(c(j)-i) = \frac{x_i}{Z(\beta)}\left(\sum_{j'} a(d_{j,j'})\right)^{\beta-1} \\
  &=& \frac{x_i}{Z(\beta)}\left(\sum_{i'} x_{i'} a(d_{i\equiv c(j),i'\equiv c(j')})\right)^{\beta-1} \equiv \frac{x_i A_i^{\beta-1}}{Z(\beta)}, \nonumber
\end{eqnarray}
where we used Eq.~(\ref{eq.pj}) and consider that $x_i \gg 1$ for all $i$.
The last equation defines the attractiveness of {\it an individual} in city $i$ as
\begin{equation}\label{eq.Ai}
A_i = \sum_{j',c(j)=i} a(d_{j,j'})= \sum_{i'} x_{i'} a(d_{i\equiv c(j),i'\equiv c(j')}).
\end{equation}
This can be thought also as the number of effective interactions available for an individual in city $i$ so that $A_i=x_i$ in the city model~(\ref{eq.a0}) and $A_i\le x_i$ otherwise (e.g., for the gravitational and exponential models). It depends only on the population $x_i$ of all cities and on the distances $d_{i,i'}$ between cities, e.g. through Eqs.~(\ref{eq.a}) or~(\ref{eq.aexp}), and therefore $A_i$ can be computed independently of the data $y_i$.  

The expected number of tokens in city $i$ is given by
\begin{equation}\label{eq.ymodel}
y_i = Y p(i) = Y \frac{x_i A_i^{\beta-1}}{Z(\beta)}.
  \end{equation}
The probability of observing $y_i$ tokens in each city of size $x_i$ is a multinomial distribution 
\begin{equation}\label{eq.multinomial}
  P(y_1, \cdots, y_N | x_1, \cdots, x_N) = Y! \prod_{i=1}^N \frac1{y_i!} \left(\frac{x_i A_i^{\beta-1}}{Z(\beta)}\right)^{y_i} \ \ .
\end{equation}
This corresponds to the likelihood $P(D|M,\theta)$  of the data $D\equiv\{y_1, \cdots, y_N\}$  -- since the populations $(x_1, \cdots, x_N)$ are fixed -- for a given model class $M$ and given parameters $\theta$. It is convenient to write the log-likelihood as
\begin{equation}\label{eq.likelihood}
\begin{alignedat}{2}
\ln P(D|M,\theta) & \equiv \ln P(y_1, \cdots, y_N | x_1, \cdots, x_N)  \\ 
&= \ln Y! - \sum_{i=1}^N \ln(y_i!) + \sum_{i=1}^N y_i \ln
 \left(\frac{x_i A_i^{\beta-1}}{Z(\beta)}\right) \ \ .
\end{alignedat}
\end{equation}

\section{Data analysis} \label{sec.analysis}

\begin{table*}[!bt]
\begin{center}
  \begin{tabular}{|c c  c | c |  c | c c |ccc|ccc|}
    \hline
    \multicolumn{3}{|c|}{Dataset} &  & \multicolumn{9}{|c|}{Models}\\\hline
        \multicolumn{3}{|c|}{} & Linear fit  & Per Capita, $\beta =1$ & \multicolumn{2}{|c|}{Cities, $\alpha=0$} & \multicolumn{3}{|c|}{Gravitational} &\multicolumn{3}{|c|}{Exponential}\\
    Country & $y$ & $N_{cities}$ & $\beta$ & $\D$ ($\Delta \D=0$) &  $\beta$   &  $\Delta \D$ & $\alpha$ &  $\beta $ &  $\Delta \D$ &  $\alpha$ &  $\beta$ & $\Delta \D$ \\
 \hline
  USA & GDP                                 & 381 & 1.11 & $25$MB& 1.12 & $-40,008$B &  $0$ & $1.12$& $-40,006$B       & $1.65$ &$1.12$ &  $-40,005$B\\
    USA & Roads                                 & 338 & 0.82 & $1.6$MB & 0.79 & $-8,358$B   & 20.4 &0.75 & $-8,596$B  &28.8 &0.77 & $-8,593$B\\
    Brazil & GDP                               & 5,480 & 1.05 & $8,309$MB& 1.17 & $-43MB$    & 14.6 & 1.24 & $-50$MB    & 17.6 & 1.21 & $-49$MB \\
   Brazil & External                                 & 5,480 & 0.97 & $0.35$MB & 1.02 & $-18$B     & 289.8 & 0.91 &  $-48$ B    & 219.4&0.93 & $-42$ B\\
   Brazil & AIDS                                 & 4,328 & 0.77 & $0.03$MB & 1.16& $-117$B     &3.1 &1.17 & $-116$B   & 4.6 &1.16 & $-118$B\\
                                                                      \hline
  \end{tabular}
  \caption{Results of the four models in five databases. $\alpha,\beta$ are the parameters in each model that best describe the data. $\D$ is the description length~(\ref{eq.dl}) of the model $M$ (measured in bytes, B) and is used to compare different models (the smaller, the better). The description length is reported (in megabytes, MB) for $M=P$ and the difference to $\D(M=P)$ is reported as $\Delta \D= \D(M) - \D(P)$ for $M=\{C,G,E\}$.  }\label{tab.results}

\end{center}
  \end{table*}

\subsection{General framework}

The models described above contain strong simplifying assumptions~\footnote{Our focus on the scaling relationship led to assumption that individuals are identical and that the token assignments are independent.} and therefore our approach here is not to test whether the data is compatible with them (we know it is not~\footnote{While in linear fitting the number of observations equals to the number of cities, our model focus on individuals $j$ and tokens of output $y$ ($X=\sum x_i, Y = \sum y_i$) so that the number of observations is much larger and the expected fluctuations (for large cities) is small. This accounts only to fluctuations of the (random) assignment of tokens and neglects fluctuations (present in the data) due to measurement imprecision and due to factors that are not part of our model.}) but instead to compare the different models. This means that instead of the {\it likelihood} $P(D|M,\theta)$ that models generate the data $D=\{y_1,\cdots,y_N\}$, computed in the previous section, we should focus on what the data $D$ tells us about the model class $M \in \{P,C,G,E\}$ and their parameters $\theta=\{\alpha,\beta\}$. This is done based on the ({\it posterior}) probability
\begin{equation}\label{eq.posterior}
P(M,\theta | D) = P(D | M,\theta) \frac{P(M,\theta)}{P(D)},
\end{equation}
computed from the three terms in the right hand side:

\begin{itemize}
\item $P(D)$ depends only on the data, act as a normalization, and does not affect the choice between models.
 \item $P(M,\theta)$ is the prior probability and is taken flat so that no {\it a priori} preference is given to any model. Specifically, we write $P(M,\theta) = P(\theta | M) P(M)$ with $P(M)=1/4$ and constant $P(\theta|M)$  in $0 \le \beta \le 2$ and $0 \le \alpha \le \alpha_{max}$, where $\alpha_{max}$ is an arbitrary maximum distance (we use $\alpha_{max}=6,371km$, Earth's radius)\footnote{This implies that $P(\theta|M)$ for our the models $P,C,G,E$ are $1,1/2,1/2\alpha_{max},1/2\alpha_{max}$, respectively.}.
  \item $P(D|M,\theta)$ is the likelihood and is evaluated numerically from Eq.~(\ref{eq.likelihood}). This is facilitated by two observations: (i) the two first terms in the log-likelihood~(\ref{eq.likelihood}) are independent of the models so that the variation across $M$ and $\theta$ depends only on the last term; (ii) in this last relevant term, the parameter $\alpha$ enters only in $A_i$ through the dependence on $a(d)$ so that for a fixed $\alpha$ the dependence of the matrix $d_{i,i'}$ is reduced to the vector $A_i$. It is thus computationally more efficient to fix $\alpha$, compute $A_i$ once, and then consider variations in $\beta$.
 
\end{itemize}

\subsection{Estimation of parameters $\theta=\{\alpha,\beta\}$}

The best parameters $\theta=\{\alpha,\beta\}$ of a given model~$M$ are the ones that maximize the posterior~$P(\theta|D,M)$. Since the priors are constant, this is equivalent to the maximization of the log likelihood~(\ref{eq.likelihood}) in the space of admissible parameters set by the priors. 

\subsection{Model selection}

In the comparison of the different model classes $M$ we account for the fact that models have different (number of) parameters $\theta$ by computing $P(M|D)$, or equivalently, the description length
\begin{equation}\label{eq.dl}
\D = -\log(P(M,D)),
\end{equation}
by integrating over all parameters $\theta$ of model $M$
\begin{eqnarray*}
  P(M,D) &= P(M|D)P(D) = \int  P(D , M,\theta) d\theta\\
 & = \int P(D | M,\theta) P(\theta|M)P(M) d\theta.
\end{eqnarray*}
The description length $\D$ corresponds to the size (in number of {\it bits}, for based 2 logarithm) of the optimal encoding of data and model~\cite{Gruenwald}. Since the priors $P(\theta|M)$ and $P(M)$ are constant, the crucial computational step is the integration of the likelihood over the parameters $\theta$. When the number of observations is $Y=\sum_i y_i$ is large (often the case for relevant urban scaling analysis), the likelihood is expected to be sharply peaked around the maximum-likelihood parameters $\theta$. In this case, the description length $\D$ is dominated by the maximum log-likelihood and further approximations can be used to compute $\D$ (e.g., the Bayesian Information Criteria). However, one should be careful using these approximations to compare non-nested models (e.g., $G$ and $E$) and around parameters $\theta$ in which the priors are discontinuous (as in the relevant case of $\alpha=0$).

\section{Results}\label{sec.results}

\begin{figure*}[!tb]
\includegraphics[width=0.65\columnwidth]{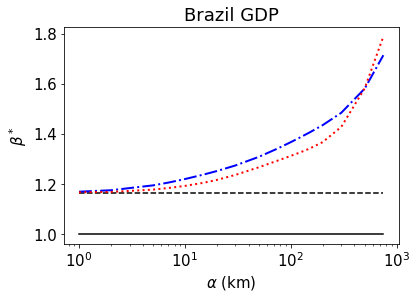}
\includegraphics[width=0.65\columnwidth]{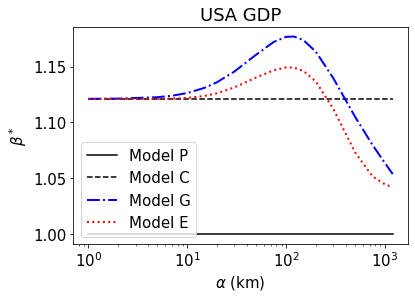}  \includegraphics[width=0.65\columnwidth]{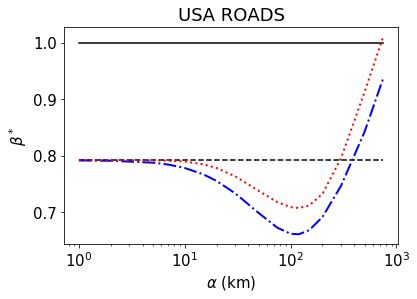}\\
  \includegraphics[width=0.65\columnwidth]{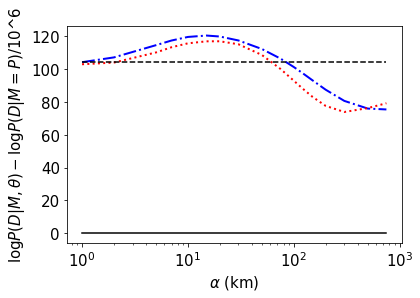}
  \includegraphics[width=0.65\columnwidth]{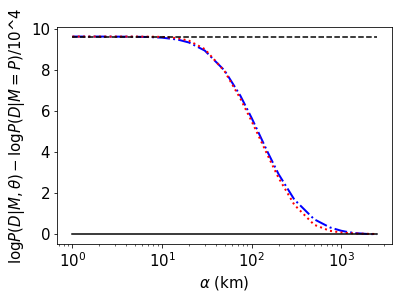}
  \includegraphics[width=0.65\columnwidth]{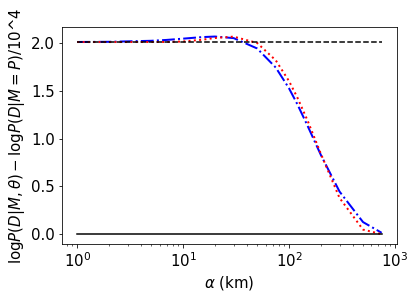}
\caption{Spatial dependence affects the choice of the scaling parameter $\beta$.  The value of $\alpha$ (measured in kilometres) is varied and the most likely value of $\beta$ is estimated for each $\alpha$. The three top panels shows the value of $\beta$ and the three bottom panels indicate the likelihood~$\mathcal{L}$ of the different models $M$ indicated in the legend. The panels on the left correspond to  GDP data from Brazil, the best model is $M=G$ with $\alpha=18$ and $\beta=1.21$. The panels on the centre correspond to GDP data from USA, the best model is the city model $M=C$ (obtained for $\alpha=0$) with $\beta=1.12$. The panels on the right correspond to  data from road miles in the USA, the best model (largest likelihood) is $M=G$ with $\alpha=24$ and $\beta=0.74$.}
\label{fig.results}
\end{figure*}

\subsection{Data}

We apply the models and data-analysis methods described above to five datasets from two different countries. For Brazil, the data on three observables $y$ (GDP, death due to external reasons, and cases of AIDS) are given for thousands of municipalities (administrative boundaries). For USA, the data on two observables $y$ (GDP, miles of roads) are given for hundreds of metropolitan areas. The USA cases can be considered as {\it the} paradigmatic examples of super- and sub-linear urban scaling laws~\cite{Bettencourt2013s}. In both countries, the average distance between two urban units is of thousands of $km$. The results of our analysis are reported in Tab.~\ref{tab.results}. The data and codes used in this paper is available in Ref.~\cite{github}.

\subsection{The effect of $\alpha$}

We start investigating the central question of this paper: does spatial proximity between cities help to explain the observations~$y$ studied in urban scaling? And, if so, does it affect the scaling exponent~$\beta$? The results in Fig.~\ref{fig.results} demonstrate that the answer to both of these questions is positive in most (but not all) cases. The top row of the figure shows that the value of the (maximum likelihood) exponent $\beta$ for a fixed $\alpha$ changes significantly with $\alpha$. The bottom row shows that often (Brazil GDP, USA Roads, but not in USA GDP) the best model is observed for $\alpha>0$. In these cases, there is an interval in $\alpha$ for which the model with geographic distance has a larger likelihood than the $\alpha=0$ case, compatible with the idea that the spatial scales we are accounting in this interval are meaningful (i.e., distances of $10-100km$ that are relevant to interacting people).

The addition of spatial interactions does not trivialize the urban scaling law, differently from the effect of city boundaries reported in Ref.~\cite{Arcaute2015}. In fact, the non-linear  scaling exponent $\beta$ is often enhanced by the spatial relation $\alpha$, i.e., super- (sub-) linear scalings $\beta>1$  ($\beta<1$) in the usual approach (at $\alpha=0$) show an even larger (smaller) value of $\beta$ for the maximum-likelihood value of $\alpha$. For instance, for Brazil GDP the estimation of $\beta$ in the non-spatial models are $1.05$ (linear fitting) and $1.17$ (city model) while in the spatial models it is $1.24$ (gravitational model)  and $1.21$ (exponential model). The same effect is observed in the case of sublinear scaling in the data for USA Roads Lengths, see Tab.~\ref{tab.results}.

\subsection{Comparing different models}

 In all our five datasets the models with non-linear scaling (C,G, and E, for which $\beta \neq 1$) are preferred over the per-capita (P) model (negative $\Delta \D$ in Tab.~\ref{tab.results}). In four of the five datasets, the models with spatial interactions ($\alpha \neq 0$ in the G and E models) are preferred over the one (C-model) that ignores it. The exception is the case of USA GDP, for which the estimated value of $\alpha$ is zero for the G model and very small ($1.65 km$) for the E model. The description length $\D$ of the C model is smaller than the one in the $G,E$ models by $2$ and $3$ bytes, respectively, indicating that the largest likelihood of the data obtained with $\alpha=1.65$ in the E model is not sufficient to justify its increased model complexity.

The comparison of the Gravitational and Exponential models reveal that both show a very similar behaviour as a function of $\alpha$ (Fig.~\ref{fig.results}), similar inferred model parameters $\alpha$ and $\beta$, and similar description lengths $\D$ (Tab.~\ref{tab.results}). This indicates that the conclusions are not very sensitive to the functional of $a(d)$,  used to account for spatial interactions (as long as they satisfy the natural constraints we used to propose $a(d)$). The most important distinction we found is between models that ignore spatial interactions (linear fitting, C model, and $\alpha=0$) and those that account for it ($\alpha > 0$ in the G and E models).

\begin{figure}[!tb]
\includegraphics[width=1\columnwidth]{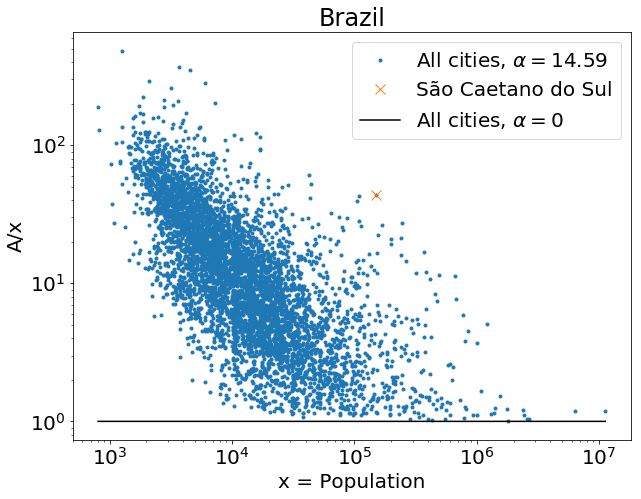}
\caption{Accounting for spatial interactions increase the attractiveness of individuals in small cities. The attractiveness $A_i$ in Eq.~(\ref{eq.Ai}) divided by the population $x_i$ is shown as a function of $x_i$ for the different Brazilian municipalities $i$. The horizontal black line correspond to the case in which spatial interactions are ignored ($\alpha=0 \Rightarrow  A_i=x_i$). The dots correspond to the result of the Gravitational model with the maximum likelihood parameters obtained for the case of GDP (see Tab.~\ref{tab.results}). }
\label{fig.a}
\end{figure}

\subsection{Increased interactivity}

We now investigate how the spatial models introduced here change the number of effective interactions of individuals. In the introduction we discussed how the GDP of cities close to large urban areas were underestimated. Our analysis reveals that spatial interactions were not a strong factor in the USA GDP data overall. This was different for Brazil GDP, when the best model is the Gravitational model with $\alpha=14.6$. For these parameters, in Fig.~\ref{fig.a} we show the increased attractiveness -- or number of interactions, $A_i$ in Eq.~(\ref{eq.Ai}) -- that individuals in different cities in Brazil experience. It fluctuates significantly from city to city because it is an intricate function of the location of all cities, but it is clear that smaller cities are more affected than larger cities.

For the case of ``S\~ao Caetano do Sul'', the attractiveness of the inhabitants of this municipality is $43.6$ times larger than assuming that interactions occur only within the municipality (i.e., $A=43.6 x$ for $\alpha=14.6$ in $M=G$). The GDP of this city is $11.0$ BR\$ (Billion reais), much larger than the per-capita expectation of $2.1$ BR\$. The city model improves this expectation to $2.7$ BR\$, still too low but better than the linear-fit estimation $1.6$ BR\$. The best spatial model (G model with $\alpha=14.6$ and $\beta=1.24$) improves the prediction to $4.5$ BR\$. Therefore, we conclude that spatial interactions can explain a considerable amount of the GDP of this municipality, even more than the inclusion of the  non-linear scaling ($\beta>0$), but that other factors remain significant. 


\section{Conclusions}
We introduced models of urban scaling laws that account for spatial interactions between individuals in different locations and that allow for rigorous statistical inference and model comparison. Results in five databases reveal that spatial interactions between cities leads to improved models and change the estimation of the urban scaling parameter $\beta$. Our approach shows how the problem~\cite{Louf2014,Arcaute2015} of the effect of the definition of the urban unit on scaling laws can be solved by including spatial interactions between different locations explicitly in the model and inference.

The framework introduced in this paper can be extended to account for more sophisticated models (of interactions), beyond the four simple models introduced here. This could include more detailed information about the proximity and connectivity between different urban areas (e.g., commuting time) and incorporate ideas proposed in models of scaling laws~\cite{Ribeiro2017}, in models of the growth of cities~\cite{BattyBook,Barthelemy}, and in methods to define boundaries of urban regions~\cite{Rozenfeld2011,Arcaute2015}.
It would be interesting to use these models to explore datasets at different spatial resolutions (e.g., neighbourhoods) and when additional information on the population in each location is available. 
The crucial point is that additional parameters and models for interactions should be inferred from the data together with the parameter $\beta$ of the urban scaling law, avoiding arbitrary choices and leaving to the data and model-comparison techniques the choice between different approaches.

\section*{Acknowledgements}
Somwrita Sarkar and Elsa Arcaute contributed with stimulating discussions.



\begin{thebibliography}{10}

\bibitem{github} The data and codes used in this paper are available at: http://www.github.com/edugalt/scaling


  
    \bibitem{Krugman2001}
      Masahisa Fujitsa, Paul R. Krugman, and Anthony J. Venables.
      The Spatial Economy: Cities, Regions, and International Trade.
      MIT Press, 2001.

      
\bibitem{BattyBook}
Michael Batty.
 {\em The new science of cities}.
 Mit Press, 2013.

\bibitem{Barthelemy}
  Marc Barthelemy.
  {\em The Structure and Dynamcis of Cities}.
  Cambridge University Press, 2016.

\bibitem{Rybski2019}  Diego Rybski, Elsa Arcaute, and Michael Batty.
  Urban Scaling Laws
  {\it Environment and Planning B: Urban Analytics and City Science} 46 (9) 1605–10 (2019).



\bibitem{Rozenfeld2011}
  Hern\'an D. Rozenfeld, Diego Rybski, Xavier Gabaix, and Hernán A. Makse.
  The Area and Population of Cities: New Insights from a Different Perspective on Cities.
  {\it American Economic Review}, 101(5), 2205–25 (2011).

\bibitem{Simini2012}
  Filippo Simini, Marta C. González, Amos Maritan, and Albert-L\'aszl\'o Barab\'asi.
  A Universal Model for Mobility and Migration Patterns.
  {\it Nature} 484 (7392): 96–100 (2012).


\bibitem{Barnes2014}
  Trevor J. Barnes and Matthew W. Wilson.
  Big Data, Social Physics, and Spatial Analysis: The Early Years.
  {\it Big Data \& Society} 1 (1): 2053951714535365 (2014).

  
  
  \bibitem{Sarkar2018}
    Somwrita Sarkar, Peter Phibbs, Roderick Simpson, and Sachin Wasnik.
    The Scaling of Income Distribution in Australia: Possible Relationships between Urban Allometry, City Size, and Economic Inequality.
    {\it Environment and Planning B: Urban Analytics and City Science} 45 (4): 603–22 (2018).

  

  \bibitem{Stewart1947}
    John Q. Stewart,
    Suggested Principles of ‘Social Physics’.
    {\it Science}, 106(2748): 179 (1947).


    \bibitem{Bettencourt2007a}
Lu{\'{i}}s M. A. Bettencourt, Jos{\'{e}} Lobo, Dirk Helbing, C.~Kuhnert, and
  Geoffrey~B West.
 Growth, innovation, scaling, and the pace of life in cities.
 {\em Proceedings of the National Academy of Sciences},
  104(17):7301--7306, 4 2007.

\bibitem{Bettencourt2010}
Lu{\'{i}}s M. A. Bettencourt, Jos{\'{e}} Lobo, Deborah Strumsky, and Geoffrey~B.
  West.
 Urban scaling and its deviations: Revealing the structure of wealth,
  innovation and crime across cities.
 {\em PLoS ONE}, 5(11):e13541, 11 2010.

    
 \bibitem{Bettencourt2013s}
Lu{\'\i}s~M. A. Bettencourt.
 The origins of scaling in cities.
 {\em Science}, 340(6139):1438--1441, 2013.

  
\bibitem{Louf2014}
R\'emi Louf and Marc Barthelemy.
 Scaling: lost in the smog.
 {\em Environment and Planning B: Planning and Design},
  41(5):767--769, 10 2014.

\bibitem{Arcaute2015}
Elsa Arcaute, Erez Hatna, Peter Ferguson, Hyejin Youn, Anders Johansson, and
  Michael Batty.
 Constructing cities , deconstructing scaling laws.
 {\em Journal of The Royal Society Interface}, (i):3--6, 2015.




\bibitem{Ribeiro2017}
  Fabiano L.  Ribeiro, Joao Meirelles, Fernando F. Ferreira, and Camilo R. Neto.
  A Model of Urban Scaling Laws Based on Distance Dependent Interactions.
  Royal Society Open Science 4 (3): 160926 (2017).

 

\bibitem{Shalizi2011}
Cosma R. Shalizi.
 Scaling and hierarchy in urban economies.
 {\em arXiv:1102.4101},  2011.


\bibitem{Leitao2016}
  Jorge C. Leitao, Jose M. Miotto, Martin Gerlach, and Eduardo G. Altmann,
Is this scaling nonlinear?
{\it Royal Society Open Science} 3, 150649 (2016) 

\bibitem{Finance2019} Olivier  Finance and Cl\'ementine Cottineau.
  Are the Absent Always Wrong? Dealing with Zero Values in Urban Scaling.
  {\it Environment and Planning B: Urban Analytics and City Science} 46(9): 1663–77 (2019).



  
\bibitem{Clauset2009}
  Aaron Clauset, Cosma R. Shalizi, and Mark E. J. Newman,
Power-Law Distributions in Empirical Data,
  SIAM Rev. 51, 661 (2009).  
  
\bibitem{Gerlach2019} Martin Gerlach and Eduardo G. Altmann.
Testing statistical laws in complex systems,
{\it Phys. Rev. Lett.} 122, 168301 (2019).
  
\bibitem{Corral2020} \'Alvaro Corral, Frederic Udina, and Elsa Arcaute.
  Truncated Lognormal Distributions and Scaling in the Size of Naturally Defined Population Clusters.
  {\it Physical Review E} 101(4): 042312 (2020).


\bibitem{Gruenwald}
  Peter D. Gr\"unwald.
  The Minimum Description Length Principle. MIT Press, 2007.

  
  
\end{thebibliography}

\end{document}